\documentclass[%
 reprint,
superscriptaddress,
 amsmath,amssymb,
 aps,
pra,
]{revtex4-1}

\usepackage{graphicx}
\usepackage{dcolumn}
\usepackage{bm}
\usepackage{epstopdf}
\usepackage{physics}

\begin{document}

\preprint{APS/123-QED}

\title{High-Fidelity, Single-Shot, Quantum-Logic-Assisted Readout in a Mixed-Species Ion Chain}

\author{C. D. Bruzewicz}
\email{colin.bruzewicz@ll.mit.edu}
\affiliation{Lincoln Laboratory, Massachusetts Institute of Technology, Lexington, Massachusetts 02421, USA}
\author{R. McConnell}
\affiliation{Lincoln Laboratory, Massachusetts Institute of Technology, Lexington, Massachusetts 02421, USA}
\author{J. A. Sedlacek }
\affiliation{Lincoln Laboratory, Massachusetts Institute of Technology, Lexington, Massachusetts 02421, USA}
\author{J. Stuart}
\affiliation{Lincoln Laboratory, Massachusetts Institute of Technology, Lexington, Massachusetts 02421, USA}
\affiliation{Center for Ultracold Atoms, Department of Physics, Massachusetts Institute of Technology, Cambridge, Massachusetts 02139, USA}
\author{W. Loh}
\affiliation{Lincoln Laboratory, Massachusetts Institute of Technology, Lexington, Massachusetts 02421, USA}
\author{J. M. Sage}
\affiliation{Lincoln Laboratory, Massachusetts Institute of Technology, Lexington, Massachusetts 02421, USA}
\author{J. Chiaverini}
\affiliation{Lincoln Laboratory, Massachusetts Institute of Technology, Lexington, Massachusetts 02421, USA}

\date{\today}

\begin{abstract}
We use a co-trapped ion ($^{88}\mathrm{Sr}^{+}$) to sympathetically cool and measure the quantum state populations of a memory-qubit ion of a different atomic species ($^{40}\mathrm{Ca}^{+}$) in a cryogenic, surface-electrode ion trap. Due in part to the low motional heating rate demonstrated here, the state populations of the memory ion can be transferred to the auxiliary ion by using the shared motion as a quantum state bus and measured with an average accuracy of 96(1)\%. This scheme can be used in quantum information processors to reduce photon-scattering-induced error in unmeasured memory qubits.

\end{abstract}

\pacs{Valid PACS appear here}

\maketitle

\section{Introduction}
Trapped atomic ions are promising candidates for the qubits for large-scale quantum processors.  High-fidelity measurement and control of systems consisting of several ions have already been demonstrated. As the number of qubits in quantum processors continues to grow, however, additional challenges will need to be addressed. For example, the fidelity of two-qubit operations is already limited by motional heating in several state-of-the-art trapped-ion experiments~\cite{tan2015multi,PhysRevLett.117.060504}, but scalable quantum processor architectures will likely require that the ions be trapped close to the trap electrode surfaces, where the motional heating rate is known to be high~\cite{PhysRevLett.97.103007,brownnutt2015ion}. Additionally, trapped-ion quantum state measurement typically involves the scattering of photons, which can cause significant computational error if these photons are absorbed by nearby unmeasured qubits.

The challenges presented by motional heating and scattering-induced error can both be mitigated with the inclusion of an auxiliary ion of another atomic species. The Coulomb interaction couples the motion of the ions, and the shared motion can be sympathetically laser cooled by addressing  only the auxiliary species~\cite{rohde2001sympathetic,barrett2003sympathetic,home2009memory}. As the cooling lasers are very far detuned from all transitions in the memory ions, the effects of motional heating are eliminated without compromising memory qubit coherence. The auxiliary ion can also be used for quantum state measurement; by first mapping the memory ion's quantum state populations to the auxiliary ion's internal states, detection can similarly proceed using light far detuned from all memory ion transitions.

Similar techniques have been implemented in previous demonstrations. For example, single-shot quantum-logic-assisted schemes have been used to transfer state population between ions, but with lower measurement accuracies~\cite{schmidt2005spectroscopy,inlek2017multi}. In other work, adaptive quantum nondemolition measurements have been used to achieve very high readout fidelities, though these measurements require several time-consuming detection cycles that would slow the operation of a quantum processor~\cite{hume2007high}. In addition, they take advantage of a unique energy level structure that is not readily extensible to the ion species typically used for quantum processing. An alternate scheme using auxiliary ions of the same species has been employed~\cite{roos2004control}, but this method adds undesirable computational overhead, as single-qubit operations are required to shelve and subsequently deshelve unmeasured memory ion state amplitudes in non-resonant levels before and after detection. State-readout based on sophisticated inter-species conditional-NOT gates has also been demonstrated elsewhere~\cite{tan2015multi}. As we show here, however, comparable measurement accuracy can be achieved with a more straightforward quantum-logic-assisted scheme.  

Using single-shot, quantum-logic-assisted readout, we measure the quantum state populations of a calcium ion with an average accuracy of 96(1)\% by monitoring the state-dependent fluorescence of an auxiliary strontium ion in a cryogenic, surface-electrode trap. We also determine the motional heating rates of the axial modes following sympathetic cooling and find them to be sufficiently low to support high-fidelity inter-species quantum logic. The high performance of these techniques permits shifting the necessary tasks of cooling and state measurement to a second species, which can reduce control error, crosstalk, and decoherence in a trapped-ion quantum processor.

\section{Experiment}
\begin{figure}
\begin{center}
\includegraphics[width=\columnwidth]{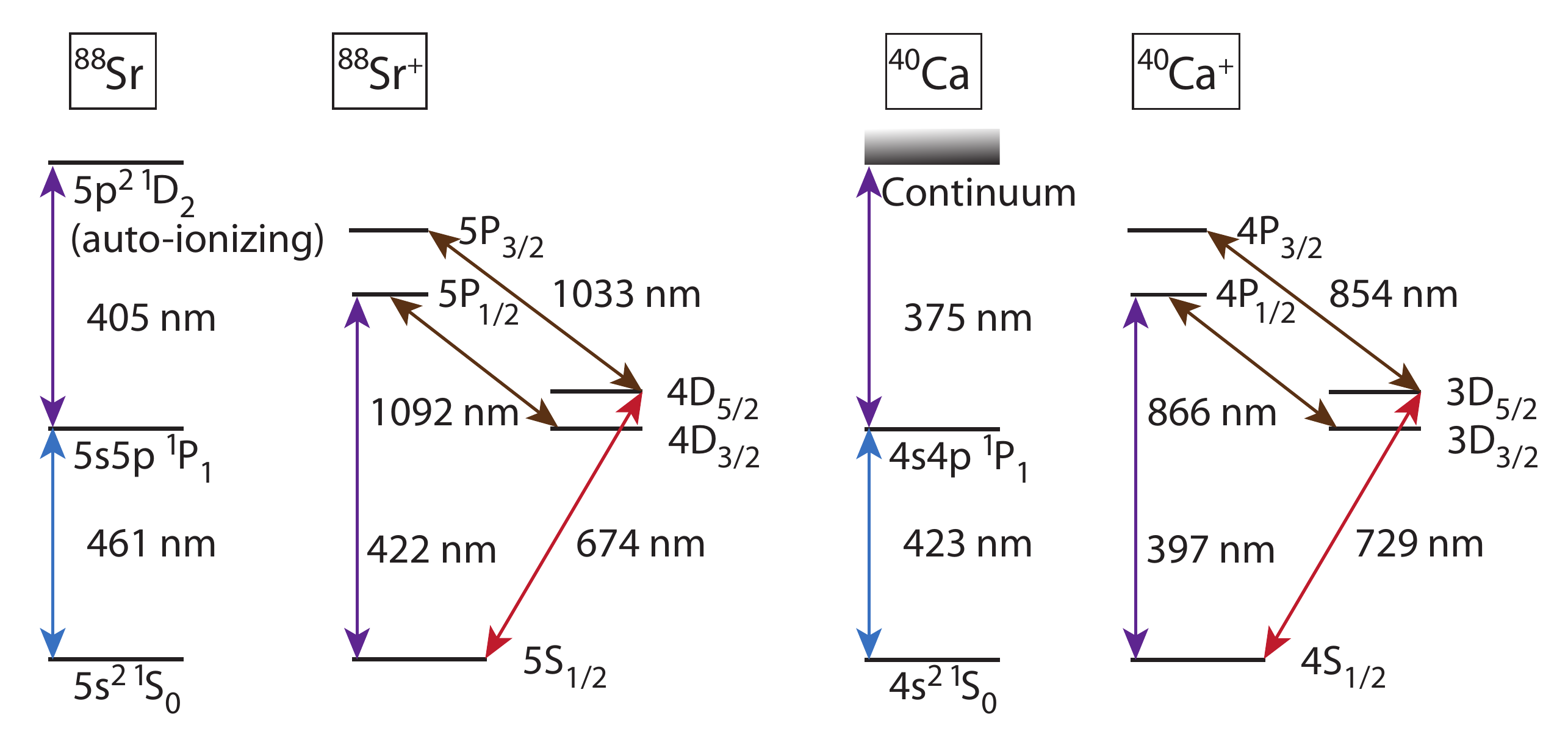}
\caption{(Color online) Relevant energy levels for cooling and manipulation of neutral and singly-ionized $^{88}\mathrm{Sr}$ and $^{40}\mathrm{Ca}$.}
\label{fig:expt}
\end{center}
\end{figure}

Both ion species ($^{40}\mathrm{Ca}^{+}$  and $^{88}\mathrm{Sr}^{+}$) are loaded from pre-cooled beams of neutral atoms. The atomic beams are generated from independent, spatially-overlapped, two-dimensional magneto-optical traps (2D-MOT) that are separated from the ion trapping chamber by a narrow differential pumping tube~\cite{bruzewicz2015scalable}. In the trapping chamber, the ion species are photoionized individually by resonant two-step processes shown schematically in Fig.~\ref{fig:expt}. The ions are confined and subsequently Doppler cooled approximately 50~$\mu$m from a segmented, linear, surface-electrode Paul trap made from photolithographically patterned niobium on a sapphire substrate. The trap chip is wirebonded to a ceramic pin grid array that is inserted into a chip carrier assembly attached to the cold head of a closed-cycle cryocooler. This cryogenic system permits operation of the ion trap at temperatures below 8~K, where motional heating rates are known to be much lower than when operated at room temperature~\cite{PhysRevLett.100.013001}. Under these conditions, the trap lifetime of the mixed-species chain is several hours.

To reduce off-resonant excitation during quantum-logic-assisted readout, the lasers driving the narrow $S_{1/2}\!\to\!D_{5/2}$ qubit transitions are filtered using high-finesse optical cavities~\cite{sterr2009ultrastable,akerman2015universal}. The small amount of light transmitted through the cavities ($\sim\!10~\mu$W) is used to injection lock diode lasers that seed tapered amplifier systems, providing sufficient laser power to drive the qubit transitions.

\section{Results}

When implementing sympathetic cooling, each experimental trial begins with a stage of Doppler cooling of all the motional modes using the strong $S_{1/2}\!\to\!P_{1/2}$ transition of the coolant ion. The in-phase and out-of-phase motion of the ion chain along the axis of the trap can then be separately cooled to the ground state using pulsed resolved sideband cooling on the narrow  $S_{1/2}\!\to\!D_{5/2}$ transition of the coolant ion. When cooling both axial modes, sideband cooling pulses addressing the different motional modes are interspersed to minimize heating of the other mode due to spontaneous emission. Although either species can be used for sympathetic cooling or memory, the longer spontaneous lifetime of the $^{40}\mathrm{Ca}^{+} 3D_{5/2}$ state ($\tau\approx1.2~\mathrm{s}$) suggests its use as the memory ion~\cite{PhysRevA.95.042507}.  Hence, for all of the measurements presented here, $^{88}\mathrm{Sr}^{+}$ is used as the sympathetic coolant.

\begin{figure}
\begin{center}
\includegraphics[width=\columnwidth]{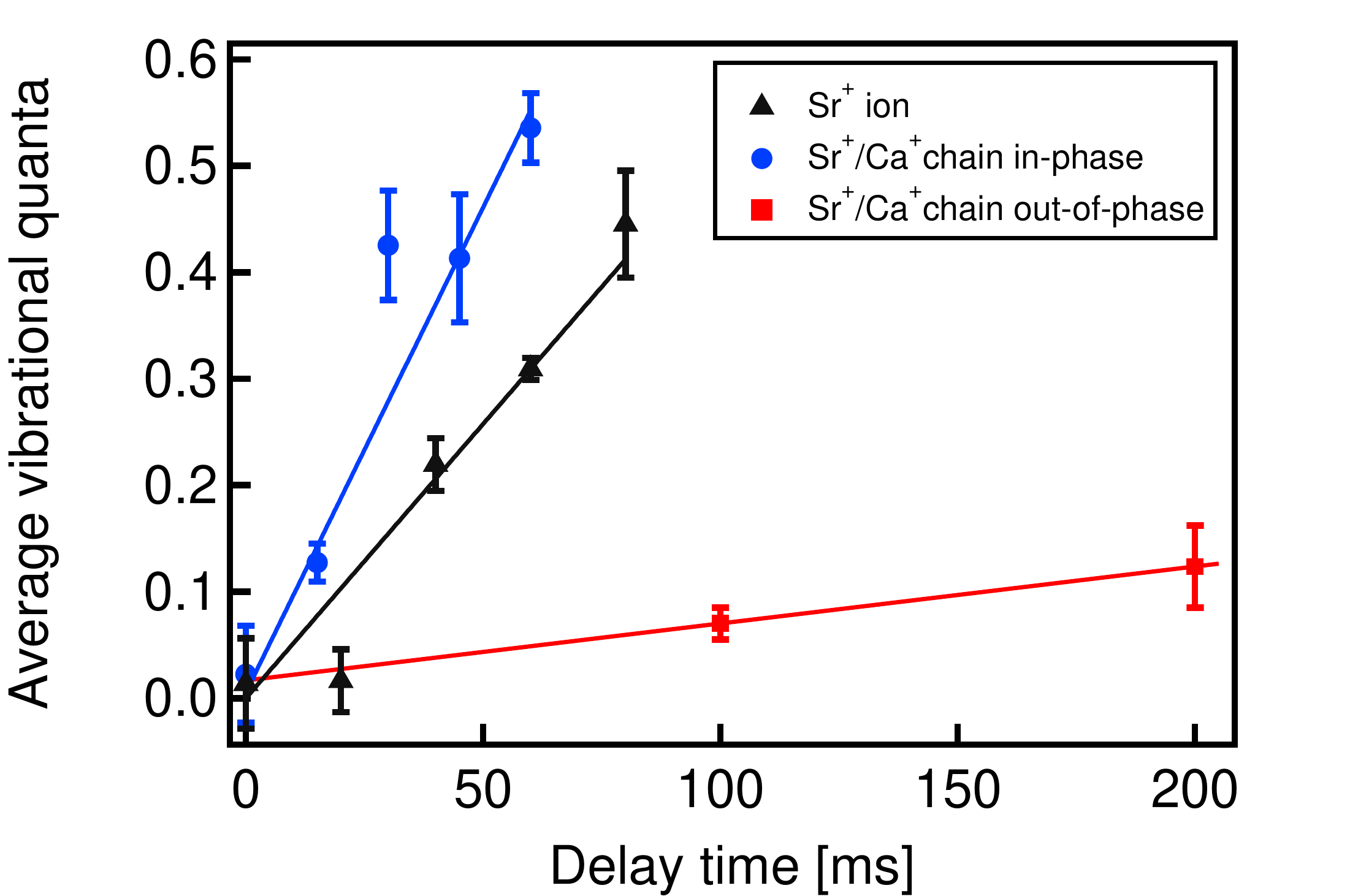}
\caption{(Color online) Axial heating rate of a single Sr$^{+}$ ion and the in-phase motion of a mixed species $\mathrm{Sr}^{+}\!/\mathrm{Ca}^{+}$ chain. In the mixed-species experiments, sideband amplitudes were measured by addressing the Sr$^{+}$ ion. Solid lines are linear fits to the measured data; extracted heating rates are given in the main text. Error bars reflect uncertainty in the extracted value of the average occupation $\bar{n}$ with quantum projection noise propagated throughout the fitting procedure.}
\label{fig:heating}
\end{center}
\end{figure}

After cooling to the ground state, motional heating rates are determined by measuring the average occupation $\bar{n}$  of a particular mode after a variable delay time using the sideband amplitude ratio technique~\cite{PhysRevLett.75.4011}.  Figure~\ref{fig:heating} shows the measured motional heating rates for a single $^{88}\mathrm{Sr}^{+}$ ion as well as both axial modes of a mixed-species chain. The heating rate of a single ion is measured as 5.2(5)~quanta/s for a 1.2~MHz axial trap frequency. This value is in line with previous measurements using ion traps operated at cryogenic temperatures~\cite{chiaverini2014insensitivity,bruzewicz2015measurement}. The 1.3~MHz Sr$^{+}$/Ca$^{+}$ in-phase motional heating rate is 9.1(7)~quanta/s,  independent of  whether the out-of-phase motion is also cooled.  The observed heating rate of the in-phase motion is consistent with the predicted value calculated from the single-ion heating rate and the Sr$^{+}$/Ca$^{+}$ modal eigenvectors~\cite{wubbena2012sympathetic}, assuming a $1/f$ scaling of the electric-field noise spectral density. Based on measurements after delays of 100 and 200~ms, the heating rate of the 2.7~MHz out-of-phase mode is 0.5(4)~quanta/s. This lower heating rate reflects the suppressed sensitivity of the out-of-phase mode to spatially-uniform electric-field noise~\cite{wineland1998experimental}. High precision measurement of this heating is limited by drifts in experimental parameters, such as the magnetic field, during the long delay times necessary to observe significant red sideband amplitude.

Using the techniques of quantum logic spectroscopy~\cite{schmidt2005spectroscopy}, we have observed Rabi oscillation of the memory ion by measuring the resonant fluorescence of the auxiliary ion. In these experiments, both axial motional modes are first cooled to the ground state, and the ions are each initialized to a single Zeeman sublevel by optical pumping. The ion chain is then illuminated for a variable time on the $\ket{S}_{\mathrm{Ca}}\equiv4S_{1/2}\!\to\!\ket{D}_{\mathrm{Ca}}\equiv3D_{5/2}$ calcium carrier transition to put the memory ion in a superposition of its internal states (denoted as ``Rabi" in Fig.~3(a)). Resonant $\pi$-pulses on the first-order, red, in-phase motional sidebands $(``\mathrm{RSB}_{\pi}"$ in Fig.~3(a)) of calcium and then strontium  transfer the $\ket{D}_{\mathrm{Ca}}$ state amplitude to the
$\ket{D}_{\mathrm{Sr}}\equiv4D_{5/2}$ state using the shared motion as a quantum state bus.  The idealized evolution of the ions' electronic and motional states during the pulse sequence is given by
\begin{align}
\mathrm{Initialize}:\ket{\psi}_{0}=&\hspace{1mm}\ket{S}_{\mathrm{Ca}}\ket{S}_{\mathrm{Sr}}\ket{0} \nonumber\\
\mathrm{Ca}^{+}\hspace{1mm}\mathrm{Rabi}:\ket{\psi}_{1}=&\hspace{1mm}\big(\alpha\ket{S}_{\mathrm{Ca}}+\beta\ket{D}_{\mathrm{Ca}}\big)\ket{S}_{\mathrm{Sr}}\ket{0} \nonumber\\
\mathrm{Ca}^{+}\hspace{1mm}\mathrm{RSB}_{\pi}:\ket{\psi}_{2}=&\hspace{1mm}\alpha\ket{S}_{\mathrm{Ca}}\ket{S}_{\mathrm{Sr}}\ket{0}  + \beta\ket{S}_{\mathrm{Ca}}\ket{S}_{\mathrm{Sr}}\ket{1} \nonumber\\ 
=&\hspace{1mm}\ket{S}_{\mathrm{Ca}}\ket{S}_{\mathrm{Sr}}(\alpha\ket{0}+\beta\ket{1}) \nonumber\\ 
\mathrm{Sr}^{+}\hspace{1mm}\mathrm{RSB}_{\pi}:\ket{\psi}_{f}=&\hspace{1mm}\alpha\ket{S}_{\mathrm{Ca}}\ket{S}_{\mathrm{Sr}}\ket{0} + \beta\ket{S}_{\mathrm{Ca}}\ket{D}_{\mathrm{Sr}}\ket{0} \nonumber\\ 
=&\hspace{1mm}\ket{S}_{\mathrm{Ca}}\big(\alpha\ket{S}_{\mathrm{Sr}}+\beta\ket{D}_{\mathrm{Sr}}\big)\ket{0},
\end{align}
where $\ket{S}_{\mathrm{Sr}}\equiv5S_{1/2}$ and $\ket{0}$ or $\ket{1}$ denotes the vibrational occupation of the in-phase motional mode.

\begin{figure}
\begin{center}
\includegraphics[width=\columnwidth]{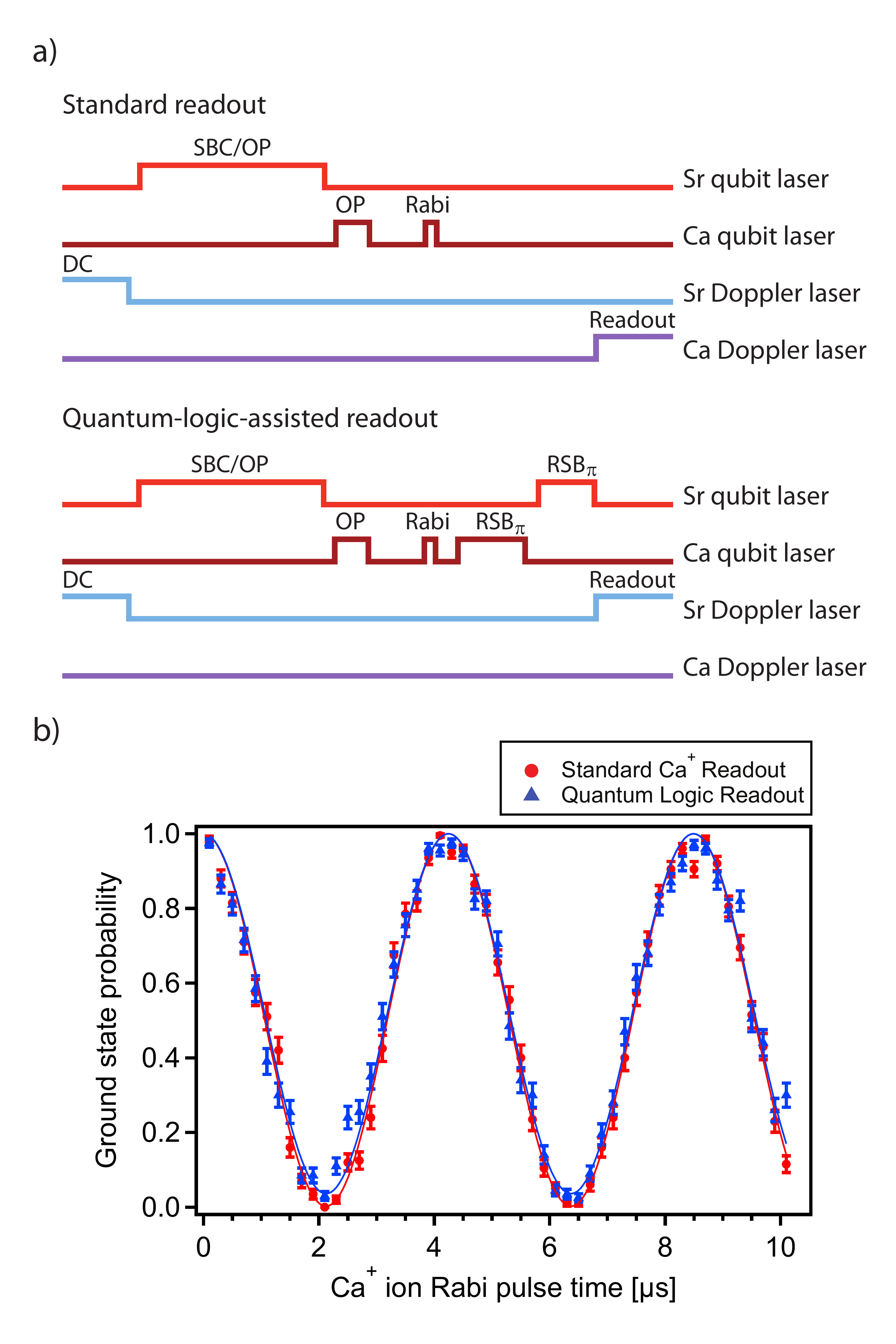}
\caption{(Color online) (a) Schematic representation of the laser pulse sequences used in standard state-dependent fluorescence detection and quantum-logic-assisted readout of calcium. The red sideband $\pi$-pulses are 59~$\mu$s and 37~$\mu$s long for Ca$^{+}$ and Sr$^{+}$, respectively. SBC: Sideband cooling of the axial motion. OP: Optical pumping to a single Zeeman sublevel of the $S_{1/2}$ state. $\mathrm{RSB}_{\pi}$: Red sideband $\pi$-pulse of the in-phase motion. DC: Doppler cooling. (b) Comparison of Rabi oscillation measured using direct calcium state detection (red) and using quantum-logic-assisted readout (blue). Solid lines are fits to Rabi oscillation lineshapes. Error bars reflect statistical uncertainty associated with quantum projection noise after 200 experimental trials per point.}
\label{fig:qls}
\end{center}
\end{figure}

The $\ket{S}_{\mathrm{Sr}}$ state population is then determined by measuring state-dependent fluorescence from the strong $5S_{1/2}\!\to\!5P_{1/2}$ strontium transition at 422~nm. By varying the length of the initial calcium carrier transition pulse, we observe high contrast Rabi oscillation that can be compared to oscillation measured using direct readout of the calcium ion state using the $4S_{1/2}\!\to\!4P_{1/2}$ calcium transition at 397~nm without implementing the sequence of red-sideband $\pi$-pulses. As shown in Fig.~\ref{fig:qls}, the quantum-logic-assisted readout data closely follow the Rabi oscillation data gathered using standard state-dependent fluorescence.

Fitting these data to Rabi oscillation lineshapes yields an average overall single-shot quantum state population transfer and readout efficiency of 96(1)\%. The discrepancy between the two readout techniques suggests that the red sideband pulses used for quantum logic are not true $\pi$-pulses and that the excited state amplitude of the memory ion is not fully transferred to the auxiliary readout ion. The pulse infidelity is likely dominated by frequency errors caused by magnetic field fluctuations as well as AC Stark shifts  induced by off-resonant coupling to nearby carrier transitions~\cite{haffner2003precision}. The measured motional heating rate contributes negligible error to the quantum state transfer and readout.

Composite pulse sequences could be used to increase the red sideband $\pi$-pulse fidelities in this scheme, but the longer pulse time required to implement such sequences puts additional requirements on the coherence of the lasers used to drive the transitions~\cite{wimperis1994broadband,cummins2003tackling}. Even if the sideband $\pi$-pulses are made perfectly accurate, the state population readout accuracy of this scheme will still be limited by the ability to completely cool the shared ion motion to the ground state, as the red sideband pulses will drive transitions from  higher-lying motional states, if occupied.   However, measurement error of the magnitude demonstrated here may still be made compatible with fault-tolerant quantum computation, if the computational gate errors remain below the error-correction threshold~\cite{divincenzo2000physical,PhysRevLett.94.010501}. For example, high-fidelity conditional logic gates between the memory qubit and $M$ additional ancilla ions of the same species can be used to encode the memory qubit state into a larger entangled state given by 
\begin{align}
\ket{\psi}=&\hspace{1mm}\alpha\ket{S}_{\mathrm{Ca}}\ket{S}_{\mathrm{Ca}^{(1)}}\ket{S}_{\mathrm{Ca}^{(2)}}\cdots\ket{S}_{\mathrm{Ca}^{(M)}}\hspace{1mm}+ \nonumber \\
&\hspace{1mm}\beta\ket{D}_{\mathrm{Ca}}\ket{D}_{\mathrm{Ca}^{(1)}}\ket{D}_{\mathrm{Ca}^{(2)}}\cdots\ket{D}_{\mathrm{Ca}^{(M)}}.
\end{align}
The calcium state populations can then be mapped to the internal states of $M+1$ strontium ions by implementing the quantum logic sideband pulse sequence for each of the $M+1$ mixed-species ion pairs. Measuring the strontium ions' states and accepting the majority outcome would reduce the overall measurement error, despite the imperfect population transfer demonstrated here. These steps can be performed in parallel, yielding improved readout accuracy without significant additional time overhead.

Quantum-logic-assisted readout offers significant advantages over single-species operation by reducing the errors accrued by unmeasured memory qubits due to incidental resonant photon scatter. Assuming isotropic emission from a fluorescing ion, the absorption probability of a neighboring ion per emitted photon $P$ is given by the ratio of the absorption cross-section to the surface area of a sphere with radius equal to the ion separation~\cite{beterov2015rydberg}. The total photon scattering-induced error probability during state readout $P_{tot}$ increases monotonically with the number of emitted photons, but in order to differentiate between the fluorescing and non-fluorescing states with the desired accuracy, a sufficient number of photons must be detected.  For total photon collection efficiency $\eta_{coll}$ and $N$ emitted fluorescence photons, this state-discrimination error probability can be made very small ($<\!10^{-4}$) for even relatively small numbers of detected photons ($n_{d}\!=\!\eta_{coll}N\!\geq\!10$)~\cite{wineland1998experimental}. 

\begin{figure}
\begin{center}
\includegraphics[width=\columnwidth]{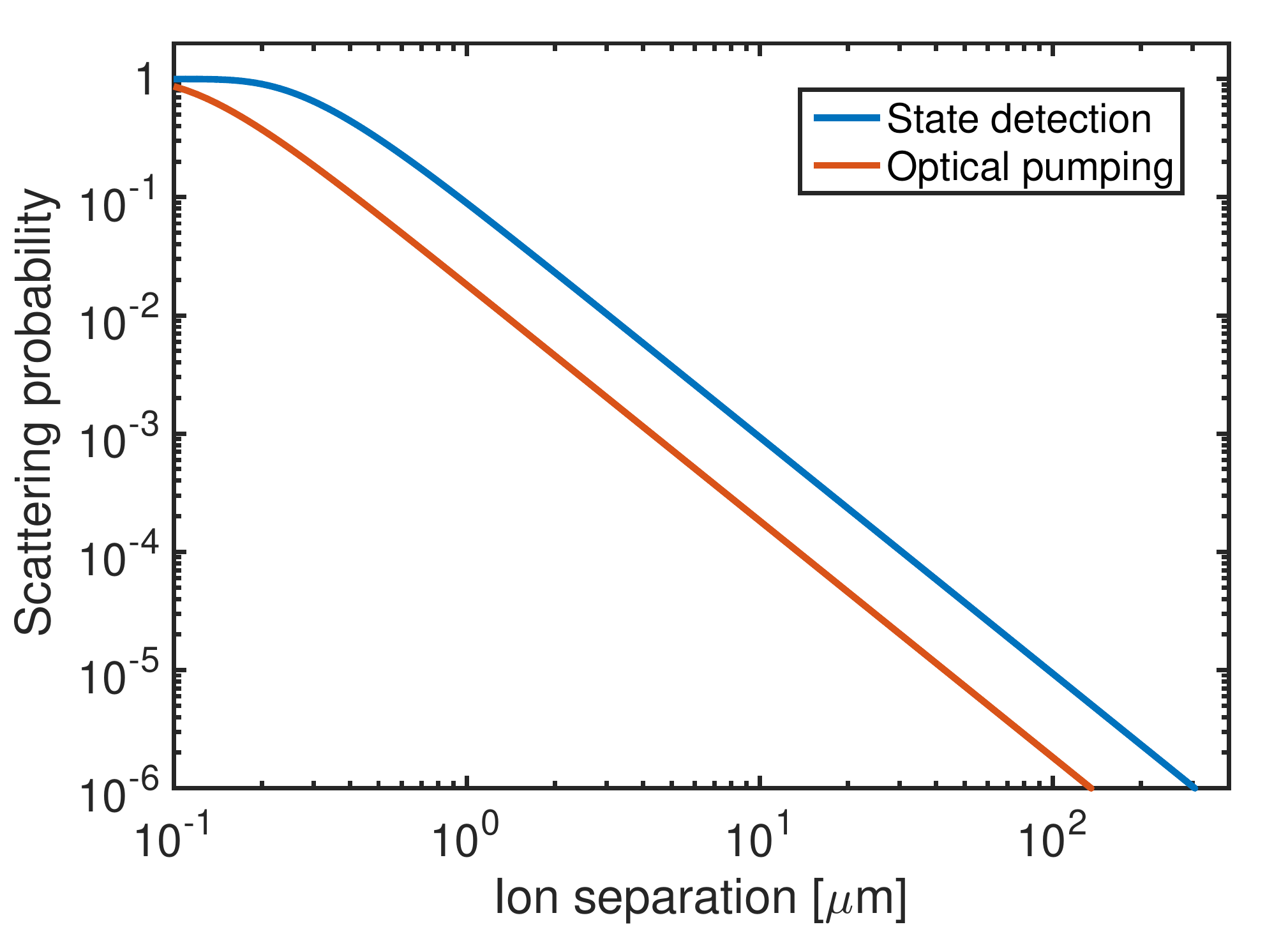}
\caption{(Color online) Calculated photon scattering probabilities due to spontaneous emission for a $\mathrm{Ca}^{+}$ ion during state detection (blue) and optical pumping (red) of a nearby $\mathrm{Ca}^{+}$ ion. The total scattering probabilities $P_{tot}$ are given by $1-(1-P)^{N}$, where $P$ is the scattering probability per photon emitted from the neighboring ion and $N$ is the total number of emitted photons. For state detection, the average number of detected photons is set to $n_{d}\!=\!10$ and the collection efficiency is taken as $\eta_{coll}\!=\!0.1,$ yielding an average total photon emission count $N\!=\!100$. For optical pumping, the photon scattering count is set to $N_{\mathrm{OP}}\!=\!10.$}
\label{fig:sep}
\end{center}
\end{figure}

Scattering-induced error probabilities based on the resonant cross-section of calcium and experimentally reasonable parameters ($\eta_{coll}\!=\!0.1$ and $n_{d}\!=\!10$) are shown in Fig.~\ref{fig:sep}. Based on these estimates, calcium ions must be separated by at least 30~$\mu\mathrm{m}$ during detection to achieve photon scattering error at the $10^{-4}$ level. This separation is significantly larger than the values observed in typical experiments using linear ion traps. Hence, in closely-spaced ion systems, quantum-logic-assisted readout would reduce scattering-induced memory errors during state detection by many orders of magnitude. For example, a $^{40}\mathrm{Ca}^{+}$ ion located $5~\mu\mathrm{m}$ from a $^{88}\mathrm{Sr}^{+}$ ion would incur an error probability of less than $10^{-13}$.

Although this readout technique can significantly reduce scattering-induced error during state detection, potentially destructive spontaneous emission still occurs during the optical pumping process that initializes the memory ion in a single Zeeman sublevel of the $S_{1/2}$ state. Specifically, decay from the excited $P_{3/2}$ state generates photons ($\lambda_{\mathrm{Ca}}\!=\!393~\mathrm{nm}$) resonant with nearby memory ions in the $S$ state. Fortunately, fewer photon emission events  are needed for high-fidelity state preparation $(N_{\mathrm{OP}}\!\approx\!10)$ compared to fluorescence detection $(N\!\approx\!10/\eta_{coll})$, so the error probability for a nearby ion is reduced. For typical ion separation of $5~\mu\mathrm{m}$ in a linear trap, the associated error from photon scattering during optical pumping is approximately $7\times10^{-4}$. Increasing the ion separation to $14~\mu\mathrm{m}$ reduces this error below $10^{-4}$.

In addition to the fluorescence light emitted by other ions, laser scattering from nearby optics and electrode surfaces will give rise to unwanted resonant photon flux during quantum state measurement. The error rates associated with this type of loss are in general difficult to model and dependent on the system under study. Quantum-logic-assisted readout using an auxiliary ion of another species reduces the scattering rate of the very far-detuned detection light, independent of its origin, to a negligibly small level for any experimentally feasible separation.  Scattering of photons associated with memory qubit optical pumping, however, can still lead to error. Given the lower laser powers and smaller number of scattered photons necessary for optical pumping, the predicted effects are reduced by an order of magnitude compared to those that would be caused by resonant detection light.

\section{Conclusion}
Here we have demonstrated high-fidelity quantum-logic-assisted state readout of a memory ion in a mixed-species chain with an average overall  population transfer and readout efficiency of 96(1)\%. By implementing sympathetic cooling and state-dependent fluorescence measurement using an auxiliary co-trapped species, the amount of spontaneously-emitted light resonant with memory ion transitions has been significantly reduced. This technique shows great promise for reducing scattering-induced error in the unmeasured qubits of a closely-spaced trapped-ion system. With the inclusion of additional ancilla qubits, quantum-logic-assisted readout may permit state measurement in a large ion-trap array with error rates compatible with fault-tolerant quantum processing.

\begin{acknowledgments}
We thank Michael Gutierrez for early technical contributions to the experiment, Kevin Obenland for fruitful discussions, and Peter Murphy, Chris Thoummaraj, and Karen Magoon for assistance with ion trap chip packaging. This work was sponsored by the Assistant Secretary of Defense for Research and Engineering under Air Force Contract \#FA8721-05-C-0002. Opinions, interpretations, conclusions, and recommendations are those of the authors and are not necessarily endorsed by the United States Government.
\end{acknowledgments}

\bibliography{TwoIonBib}
\end{document}